\documentclass[twocolumn]{revtex4}
\usepackage{amssymb,epsf}
\usepackage{latexsym}
\usepackage{xcolor}
\usepackage{float}
\begin{document}

\title{Generalized uncertainty principle and burning stars}
\author{H. Moradpour\footnote{h.moradpour@riaam.ac.ir}, A. H. Ziaie\footnote{ah.ziaie@maragheh.ac.ir}, N. Sadeghnezhad\footnote{nsadegh@maragheh.ac.ir}}
\address{Research Institute for Astronomy and Astrophysics of Maragha (RIAAM), University of Maragheh, P.O. Box 55136-553, Maragheh, Iran}
\date{\today}

\begin{abstract}
Gamow's theory of the implications of quantum tunneling on the
star burning has two cornerstones including quantum mechanics and
equipartition theorem. It has vastly been proposed that both of
these foundations are affected by the existence of a non-zero
minimum for length which usually appears in quantum gravity
scenarios and leads to the Generalized Uncertainty principle
(GUP). Mathematically, in the framework of quantum mechanics, the
effects of GUP are considered as perturbation terms. Here,
generalizing the de Broglie wavelength relation in the presence of
minimal length, GUP correction to the Gamow's temperature is
calculated and in parallel, an upper bound for the GUP parameter
is estimated.
\end{abstract}

\maketitle

\subsection*{Introduction}

As the first step of star burning, their constituents should
overcome the Coulomb barrier to participate in the Nuclear fusion
(NF), the backbone of star burning. It means that when the primary
gas ingredients have mass $m$ and velocity $v$, then using the
equipartition theory, one gets

\begin{eqnarray}\label{e1}
\frac{1}{2}m v^2=\frac{3}{2}K_BT\geq U_c(r_0),
\end{eqnarray}

\noindent where $K_B$ denotes the Boltzmann constant, leading to

\begin{eqnarray}\label{e2}
T\geq\frac{2Z_iZ_je^2}{3K_Br_0}\simeq 1\cdot1 \times
10^{10}\frac{Z_iZ_j}{r_0},
\end{eqnarray}

\noindent for temperature required to overcome the Coulomb
barrier. Here, $U_c(r_0)=\frac{Z_iZ_je^2}{r_0}$ denotes the
maximum of the Coulomb potential between the $i$-th and $j$-th
particles \cite{1}. Therefore, NF happens whenever the temperature
of primary gas is comparable to Eq.~(\ref{e2}) which clearly
emphasizes that in association with the heavier nuclei, NF happens
in higher temperatures. On the other hand, for the temperature of
a cloud with mass $M$ and radius $R$, we have \cite{1}

\begin{eqnarray}\label{e3}
\mathcal{T}\approx4\times10^6(\frac{M}{M_\odot})(\frac{R_\odot}{R}),
\end{eqnarray}

\noindent where $M_\odot$ and $R_\odot$ are the Sun mass and
radius, respectively. Clearly, $\mathcal{T}$ and $T$ are far from
each other meaning that NF cannot cause the star burning \cite{1}.
Therefore, we need a process that decreases Eq.~(\ref{e2}) to the
values comparable to Eq.~(\ref{e3}). Thanks to the quantum
tunnelling, overcoming the Coulomb barrier becomes possible which
finally lets star burn \cite{1}. Indeed, if the distance between
the particles becomes of the order of their de Broglie wavelength
($r_0\simeq\frac{\hbar}{p}\equiv\lambda_Q$), then the quantum
tunnelling happens and simple calculations lead to \cite{1}

\begin{eqnarray}\label{e4}
\texttt{T}\geq\frac{2Z_iZ_je^2}{3K_B\lambda_Q}\simeq 9\cdot6
\times 10^{6}Z_i^2Z_j^2(\frac{m}{\frac{1}{2}}),
\end{eqnarray}

\noindent instead of Eq.~(\ref{e2}) for the temperature required
to launch star burning. $\lambda$ has also been obtained by
solving $\frac{p^2}{2m}=U_c(r_0)\bigg|_{r_0=\lambda_Q}$ which
gives~\cite{1}

\begin{eqnarray}\label{lambdaq}
\lambda_Q=\frac{\hbar^2}{2mZ_iZ_je^2}.
\end{eqnarray}

\noindent This achievement means that the quantum tunneling provides
a platform for NF in stars \cite{1}. As an example, for Hydrogen
atoms, one can see that the quantum tunneling leads to
$\texttt{T}\simeq 9\cdot6 \times 10^{6}$ (comparable to~(\ref{e3}))
as Gomow's temperature in which NF is underway. Based on the above
argument, it is expect that any change in $p$ affects $\lambda$ and
thus the results.

It is also useful to mention here that the quantum tunneling theory
allows the above process since the tunneling probability is not
zero. Indeed, quantum tunneling is also the backbone of Gamow's
theory on $\alpha$ decay process \cite{gam}. Relying on the
inversion of Gamow formula in $\alpha$ decay theory, that gives the
transmission coefficient, a method has also been proposed for
studying the inverse problem of Hawking radiation \cite{prd}.

The backbone of quantum mechanics is the Heisenberg uncertainty
principle (HUP) written as

\begin{eqnarray}\label{HUP1}
\Delta x\Delta p\geq\frac{\hbar}{2},
\end{eqnarray}

\noindent where $x$ and $p$ are ordinary canonical coordinates
satisfying $[x_i,p_j]=i\hbar\delta_{ij}$, and modified in the
quantum scenarios of gravity as \cite{prdgup,prdgup1}

\begin{eqnarray}\label{GUP1}
(\Delta X)(\Delta
P)\geq\frac{\hbar}{2}(1+\frac{\beta_0\l_p^2}{\hbar^2}(\Delta
P)^{2}).
\end{eqnarray}

\noindent called GUP where $l_p$ denotes the Planck length, and
$\beta_0$ is the GUP parameter. Here, $X$ and $P$ denote the
generalized coordinates, and we work in a framework in which
$X_i=x_i$, and up to the first order of $\beta$, we have
$P_i=p_i(1+\frac{\beta_0\l_p^2}{3\hbar^2}p^2)$ and
$[X_i,P_j]=i\hbar(1+\frac{\beta_0\l_p^2}{\hbar^2}P^2)\delta_{ij}$
\cite{2,3}. Moreover, GUP implies that there is a non-zero minimum
for length as $(\Delta X)_{min}=\sqrt{\beta_0}~\l_p$. Indeed, the
existence of non-zero minimum for length also emerges even when
the gravitational regime is Newtonian \cite{mead1}, a common
result with quantum gravity scenarios \cite{more4}. More studies
on quantum gravity can be traced in
Refs.~\cite{more01,more02,more03,more04}. There are various
attempts to estimate the maximum possible upper bound on $\beta_0$
\cite{LowGUP1,LowGUP2,LowGUP3,LowGUP4,HighGUP1,HighGUP2,HighGUP3,HighGUP4,HighGUP5,HighGUP6,Bushev2019,Ghosh2014,Feng2017,theo2,0,01,02,03},
and among them, it seems that the maximum estimation for the upper
bound is of the order of $10^{78}$ \cite{0}. The implications of
GUP on the star evolution \cite{morad,shab}, and the
thermodynamics of various gases
\cite{3,morad2,more1,more2,more3,more4} have also attracted more
attentions.

Indeed, the existence of minimal length and gravity leads to the
emergence of GUP \cite{more4} and it affects thermodynamics
\cite{3,morad2,more1,more2,more3,more4} and quantum mechanics
\cite{prdgup,prdgup1}, as $P$ can be expand as a function of $p$.
The letter deals with the problem of the effects of GUP on star
burning launched by quantum tunnelling. Loosely speaking, we are
going to find the effects of minimal length on $\texttt{T}$.

\subsection*{GUP corrections to the tunnelling temperature}

Up to the first order of $\beta_0$, for thermal energy per
particle with temperature $T$, we have \cite{3}

\begin{eqnarray}\label{GUP2}
\langle \frac{P^2}{2m}\rangle=\langle
K\rangle=\frac{3}{2}K_BT-3\frac{\beta_0\l_p^2}{\hbar^2}m K_B^2T^2.
\end{eqnarray}

\noindent Mathematically, one should find the corresponding de
Broglie wavelength by solving equation

\begin{eqnarray}\label{GUP3}
\frac{P^2}{2m}=U_c(r_0)\bigg|_{r_0=\lambda}.
\end{eqnarray}

\noindent Inserting the result into

\begin{eqnarray}\label{GUP4}
\frac{3}{2}K_BT-3\frac{\beta_0\l_p^2}{\hbar^2}m K_B^2T^2\geq
U_c(r_0)\bigg|_{r_0=\lambda},
\end{eqnarray}

\noindent one can finally find the GUP corrected version of
Eq.~(\ref{e4}).

In order to proceed further and in the presence of the quantum
features of gravity, let us introduce the generalized de Broglie
wavelength as

\begin{eqnarray}\label{l1}
\lambda_{\rm GUP}\equiv\frac{\hbar}{P}.
\end{eqnarray}

\noindent It is obvious that as $\beta_0\rightarrow0$, one obtains
$P\rightarrow p$ and thus $\lambda_{\rm
GUP}\rightarrow\frac{\hbar}{p}$ which is the quantum mechanical
result. Indeed, up to the first order of $\beta$, we have
$\lambda_{\rm
GUP}=\lambda_Q(1-\frac{\beta_0l_p^2}{3\lambda^2_Q})$. Now,
inserting $\lambda_{\rm GUP}$ into Eq.~(\ref{GUP3}), and then
combining the results with Eq.~(\ref{GUP4}), we can find

\begin{eqnarray}\label{TGUP1}
T_{\rm GUP}^\pm=\frac{\hbar^2\left(1\pm\sqrt{1-8\beta_0l_p^2
mK_B\texttt{T}/\hbar^2}\right)}{4\beta_0 K_Bl_p^2 m}.\\\nonumber
\end{eqnarray}

\noindent in which Eq.~(\ref{e4}) has been used for
simplification. In order to have an estimation of the order of
$l_p^2 mK_B\texttt{T}/\hbar^2$, let us consider the Hydrogen atom
for which $m=\frac{1}{2}$ (here, $m$ is the reduced mass of the
Hydrogen nucleus as the primary gas constituents \cite{1}). Now,
since $l_p\propto10^{-35}$, $K_B\propto10^{-32}$,
$\hbar\propto10^{-34}$, and $\texttt{T}\propto10^6$, one easily
finds $l_p^2 mK_B\texttt{T}/\hbar^2\propto10^{-27}$. Moreover,
because the effects of GUP in the quantum mechanical regimes are
poor \cite{more4}, a reasonable and basic assumption could be that
$\beta_0l_p^2 mK_B\texttt{T}/\hbar^2\ll1$. Indeed, if
$\beta_0\ll10^{27}$, then we always have $\beta_0l_p^2
mK_B\texttt{T}/\hbar^2\ll1$ meaning that $10^{27}$ is an upper
bound for $\beta_0$ which is well comparable to previous works
\cite{LowGUP1,LowGUP2,LowGUP3,LowGUP4,HighGUP1,HighGUP2,HighGUP3,HighGUP4,HighGUP5,HighGUP6,Bushev2019,Ghosh2014,Feng2017,theo2,0,01,02,03,morad2}.
Therefore, we are allowed to expand the results.

Expanding the above solutions, and bearing in mind that the true
solution should cover $\texttt{T}$ at $\beta=0$, one can easily
find that $T_{\rm GUP}^-$ is the proper solution leading to

\begin{eqnarray}\label{TGUP2}
T_{\rm
GUP}^-=\texttt{T}\big(1+2\beta_0(\frac{l_p^2mK_B}{\hbar^2})\texttt{T}\big).
\end{eqnarray}

\noindent up to the first order of $\beta_0$. Hence, since it
seems that $\beta_0$ is positive
\cite{LowGUP1,LowGUP2,LowGUP3,LowGUP4,HighGUP1,HighGUP2,HighGUP3,HighGUP4,HighGUP5,HighGUP6,Bushev2019,Ghosh2014,Feng2017,theo2,0,01,02,03,morad2},
one can conclude that $\texttt{T}<T_{\rm GUP}^-$.

\subsection*{Conclusion}

Motivated by the GUP proposal and the vital role of HUP in quantum
mechanics and thus, quantum tunnelling letting star burn, we
studied the effects of GUP on the Gamow's temperature. In order to
continue, GUP modification to the de Broglie wavelength was
addressed which finally helped us find GUP correction to the
Gamow's temperature and also estimate an upper bound for $\beta_0$
($10^{27}$) which agrees well with previous works
\cite{LowGUP1,LowGUP2,LowGUP3,LowGUP4,HighGUP1,HighGUP2,HighGUP3,HighGUP4,HighGUP5,HighGUP6,Bushev2019,Ghosh2014,Feng2017,theo2,0,01,02,03,morad2}.

Finally and based on the obtained results, it may be expected that
GUP also affects the transmission coefficients (Gamow's formula)
\cite{gam,prd,more4} meaning that the method of Ref.~\cite{prd} will
also be affected. This is an interesting topic for future study
since Hawking radiation is a fascinating issue in black hole physics
\cite{wald}.

\subsection*{Acknowledgments}
The authors would like to appreciate the anonymous referees for
their valuable comments.



\begin{thebibliography}{99}
\bibitem{1} D. Prialnik, \textit{An Introduction to the Theory of Stellar Structure and Evolution}, (Cambridge University Press, 2000).
\bibitem{gam} G. Gamow, Z. Phys. 51, 204 (1928).
\bibitem{prd} S. H. V\"{o}lkel, R. Konoplya, K. D. Kokkotas, Phys. Rev. D 99, 104025 (2019).
\bibitem{prdgup} A. Kempf, G. Mangano, R. B. Mann, Phys. Rev. D 52, 1108 (1995).
\bibitem{prdgup1} A. Kempf, Phys. Rev. D 54, 5174 (1996).
\bibitem{2} S. Das, E. C. Vagenas. Phys. Rev. Lett. 101, 221301 (2008).
\bibitem{3} M. A. Motlaq, P. Pedram, J. Stat. Mech. P08002 (2014).
\bibitem{mead1} C. A. Mead, Phys. Rev. \textbf{135}, 849 (1964).
\bibitem{more4} S. Hossenfelder, Living Rev. Relativity. 16, 1 (2013).
\bibitem{more01} M. J. Lake et al., Class. Quant. Grav. 36, no. 15, 155012 (2019).
\bibitem{more02} M. J. Lake, M. Miller, S. D. Liang, Universe 6, 56 (2020).
\bibitem{more03} M. J. Lake, Quantum Rep. 3, 196 (2021).
\bibitem{more04} M. J. Lake, [to appear in Touring the Planck Scale, Antonio Aurilia Memorial Volume,
Springer 2022] https://arxiv.org/abs/2008.13183.
\bibitem{LowGUP1} D. Park, arXiv:2003.13856 (2020).
\bibitem{LowGUP2} P. Bosso, S. Das, I. Pikovski, M. R. Vanner, Phys. Rev. A \textbf{96}, 023849 (2017)
\bibitem{LowGUP3} I. Pikovski, M. R. Vanner, M. Aspelmeyer, M. S. Kim, C. Brukner, Nature Phys. \textbf{8}, 393 (2012)
\bibitem{LowGUP4} S. Das, R. B. Mann, Phys. Lett. B \textbf{704}, 596 (2011).
\bibitem{HighGUP1} G. G. Luciano, L. Petruzziello, Eur. Phys. J. C \textbf{79}, 283 (2019).
\bibitem{HighGUP2} G. Gecim, Y. Sucu, Phys. Lett. B \textbf{773}, 391-394 (2017).
\bibitem{HighGUP3} V. Husain, S. S. Seahra, E. J. Webster, Phys. Rev. D \textbf{88}, 024014 (2013).
\bibitem{HighGUP4} W. Chemissany, S. Das, A. F. Ali, E. C. Vagenas, JCAP \textbf{1112}, 017 (2011).
\bibitem{HighGUP5} M. Sprenger, M. Bleicher, P. Nicolini, Class. Quant. Grav. \textbf{28}, 235019 (2011).
\bibitem{HighGUP6} T. Zhu, J. R. Ren, M. F. Li, Phys. Lett. B \textbf{674}, 204 (2009).
\bibitem{Bushev2019} P. A. Bushev, J. Bourhill, M. Goryachev, N. Kukharchyk, E. Ivanov, S. Galliou, M. E. Tobar, S. Danilishin, Phys. Rev. D \textbf{100}, 066020 (2019).
\bibitem{Ghosh2014} S. Ghosh, Class. Quant. Grav. \textbf{31}, 025025 (2014).
\bibitem{Feng2017} Z. W. Feng, Sh. Z. Yang, H. L. Li, X. T. Zu, Phy. Lett. B \textbf{768}, 81-85 (2017).
\bibitem{theo2} G. G. Luciano, L. Petruzziello, Eur. Phys. J. C \textbf{79}, 283 (2019).
\bibitem{0} F. Scardigli, R. Casadio, Eur. Phys. J. C 75, 425 (2015).
\bibitem{01} F. Feleppa et al, EPL 135, 40003 (2021).
\bibitem{02} M. Mohammadi Sabet et al, Phys. Scr. 96, 125016 (2021).
\bibitem{03} S. Aghababaei et al, Phys. Scr. 96, 055303 (2021).
\bibitem{morad} H. Moradpour, A. H. Ziaie, S. Ghaffari, F. Feleppa, MNRAS 488(1), L69 (2019).
\bibitem{shab} H. Shababi, K. Ourabah, Eur. Phys. J. Plus 135, 697 (2020).
\bibitem{morad2} H. Moradpour, S. Aghababaei, A. H. Ziaie, Symmetry 13, 213 (2021).
\bibitem{more1} L. N. Chang, D. Minic, D. Okamura, T. Takeuchi, Phys. Rev. D 65, 125028 (2002).
\bibitem{more2} T. Fityo, Phys. Lett. A 372, 5872 (2008).
\bibitem{more3} P. Wang, H. Yang, X. J. Zhang, High Energy Phys. 2010, 1 ( 2010).
\bibitem{wald} R. M. Wald, Living Rev. in Rel. 4, 6 (2001).

\end{thebibliography}
\end{document}